\documentclass[preprint,aps,12pt,notitlepage,nofootinbib,tightenlines]{revtex4}
\usepackage{amsmath}
\usepackage{bm}
\usepackage{times}
\usepackage{braket}
\usepackage{color}
\usepackage{epsfig}
\usepackage{slashed}
\usepackage{hyperref}
\usepackage{multirow}
\newcommand{\beq}{\begin{eqnarray}}
\newcommand{\eeq}{\end{eqnarray}}
\newcommand{\be}{\begin{equation}\begin{aligned}}
\newcommand{\ee}{\end{aligned}\end{equation}}

\newcommand{\gev}{\text{GeV}}

\def \cpc{ {\bf Chin. Phys. C} }
\def \nima{ {\bf Nucl. Instrum. Meth. A} }

\def \ijmpa{ {\bf Int. J. Mod. Phys. A}  }
\def \copc{ {\bf Comput. Phys. Commun. } }
\def \epjc{{\bf Eur. Phys. J. C} }

\def \npb{ {\bf Nucl. Phys. B} }
\def \plb{ {\bf Phys. Lett. B} }

\def \prt{  {\bf Phys. Rept.} }
\def \prd{ {\bf Phys. Rev. D} }
\def \prl{ {\bf Phys. Rev. Lett.}  }

\def \jhep{ {\bf JHEP}  }
\def \appb{ {\bf Acta Phys. Polon. B} }
\definecolor{Red}{rgb}{1.,0.,0.}

\definecolor{Blue}{rgb}{0.,0.,1.}

\definecolor{nicered}{rgb}{0.7,0.1,0.1}
\definecolor{nicegreen}{rgb}{0.1,0.5,0.1}
\def\lsim{ {\ \lower-1.2pt\vbox{\hbox{\rlap{$<$}\lower6pt\vbox{\hbox{$\sim$}}}}\ } }
\def\gsim{ {\ \lower-1.2pt\vbox{\hbox{\rlap{$>$}\lower6pt\vbox{\hbox{$\sim$}}}}\ } }

\hypersetup{colorlinks,citecolor=nicegreen,linkcolor=nicered}
\begin{document}
\title{Probing the top-Higgs boson FCNC couplings via the $h\to \gamma\gamma$ channel at the HE-LHC and FCC-hh}
\author{Yao-Bei Liu$^{1}$\footnote{E-mail: liuyaobei@hist.edu.cn}}
\author{Stefano Moretti$^{2}$\footnote{E-mail: s.moretti@soton.ac.uk}}
\affiliation{1. Henan Institute of Science and Technology, Xinxiang 453003, P.R. China \\  
2. School of Physics \& Astronomy, University of Southampton, Highfield, Southampton SO17 1BJ, UK }

\begin{abstract}
We investigate the sensitivity of future searches for the top-Higgs boson Flavour Changing Neutral Current (FCNC) couplings $tqh$~($q= u, c$) at the proposed High Energy  Large Hadron Collider~(HE-LHC) and Future Circular Collider in hadron-hadron mode (FCC-hh). We perform a full simulation for two processes  in the $h\to \gamma\gamma$ decay channel (where $h$ is the discovered Higgs state):  single top quark FCNC production in association
with the Higgs boson (plus a jet) and top quark pair production with FCNC decays
 $t\to qh$.  All the relevant backgrounds are considered in a cut based analysis to obtain the limits on the Branching Ratios (BRs) of $t\to uh$ and $t\to ch$. It is shown that, at the HE-LHC with an integrated luminosity of 15 ab$^{-1}$ and at the FCC-hh with an integrated luminosity of 30 ab$^{-1}$, {{the BR($t\to uh$) (BR($t\to ch$)) can be probed, respectively, to $7.0~(8.5)\times 10^{-5}$ and $2.3~(3.0) \times 10^{-5}$ at the 95\% Confidence Level (CL) (assuming a 10\%  systematic  uncertainty on the background),
which is almost two orders of magnitude better than the current 13 TeV LHC experimental results.}}
\end{abstract}

\maketitle
\newpage
\section{Introduction}
The discovery of a 125 GeV Higgs boson~\cite{atlas-higgs,cms-higgs} at the Large Hadron Collider (LHC)\footnote[1]{Henceforth, it will be denoted by the symbol $h$.} was a landmark in the history of particle physics and it has opened up a new area of direct searches for Beyond the Standard Model (BSM) phenomena, since the $h$ state may well be the portal into a New Physics (NP)  world. Possible signals of NP are Flavour Changing Neutral Current (FCNC) interactions between the Higgs boson, the $t$-quark and a $u$- or $c$-quark, i.e., the vertex $tqh$ ($q=u,c$). In the SM, the FCNC top quark decays $t\to qh$~($q=u,c$) are  forbidden at the tree level and strongly suppressed  at the loop level due to the Glashow-Iliopoulos-Maiani (GIM) mechanism \cite{Glashow:1970gm}. For instance, the predicted  BR($t\to qh$)'s with $q=u,c $ are expected to be of $\mathcal{O}(10^{-12}-10^{-17})$~\cite{Mele:1998ag,AguilarSaavedra:2004wm,AguilarSaavedra:2009mx} at one-loop level and are therefore out of range for current and near future experimental sensitivity. However, in some NP models the BRs for the $t\to qh$ decays are predicted to be  in the range of $\mathcal{O}(10^{-6}-10^{-3})$~\cite{DiazCruz:2001gf,Cao:2014udj,Gao:2014lva,He:1998ie,He:2002fd,Kao:2011aa,Chen:2013qta,Han:2013sea, Abbas:2015cua,Botella:2015hoa,Arroyo:2013tna,Yang:2013lpa,Badziak:2017wxn}. Thus, any observation of such FCNC processes would be a clear signal of BSM dynamics.

Recently, the most stringent
constraint on the top-Higgs FCNC couplings through direct measurements was reported by the CMS and ATLAS collaborations~\cite{cms-8,atlas-8,atlas13-1,atlas13-2,atlas13,cms13}, by searching for $t\bar{t}$ production with one top decaying to $Wb$ and the other assumed to decay to $hq$. Corresponding to 36.1 (35.9) fb$^{-1}$  of data at the center-of-mass~(c.m.) energy of 13 TeV for ATLAS~(CMS), the 95\% Confidence Level~(CL) upper limits are summarised in
Tab.~\ref{table:top_fcnc_results}. In addition to the
direct collider measurements, indirect constrains on an anomalous $tqh$ vertex can be obtained from
the observed $D^{0}-\bar{D^{0}}$ mixing and $Z\to c\bar{c}$ decays, where the upper limits of BR$(t\to qh) < 5 \times 10^{-3}$~\cite{Aranda:2009cd} and
BR$(t\to qh) < 0.21\%$~\cite{Larios:2004mx} are obtained, respectively. From a phenomenological viewpoint, the top-Higgs FCNC interactions have been studied extensively at hadron colliders within many NP scenarios~\cite{Tait:2000sh,AguilarSaavedra:2000aj,He:1999vp,Cao:2007ea,Han:2008xb,Zhang:2010dr,Berger:2011ua,Kobakhidze:2014gqa,Atwood:2013ica,Chen:2015nta,Khatibi:2014via}.
Besides,
many phenomenological studies using model-independent methods have also been
performed via either anomalous top decay or anomalous top production processes~\cite{Wang:2012gp,zhangcen,Craig:2012vj,Shi:2019epw,Liu:2016dag,Sun:2016kek}.

A more promising result was put forward by the ATLAS Collaboration~\cite{atlas-14-3000,atlas1-14-3000}, which has predicted the sensitivities BR$(t\to uh)<2.4\times 10^{-4}$ and BR$(t\to ch)<1.5\times 10^{-4}$ at $95\%$ CL at the High Luminosity LHC~(HL-LHC). One can expect to improve further these limits at higher c.m. energies~\cite{Mandrik:2018yhe}. The future High Energy LHC~(HE-LHC) with 27 TeV c.m. energy~\cite{HE-LHC} and Future Circular Collider in hadron-hadron mode~(FCC-hh) with 100 TeV c.m. energy~\cite{FCC} have great potential to pursue direct evidence of  top-Higgs FCNC couplings with integrated luminosities of 15 ab$^{-1}$
and 30 ab$^{-1}$ in their final stages, respectively. While rather distant in the future, there are
a lot of studies in literature that have shown how these machines can greatly improve the scope of previous accelerators in pursuing BSM
 searches \cite{HELHC-1,HELHC-2,HELHC-3,HELHC-4}. So it is rather appropriate to assess their scope in accessing $tqh$ vertices too, the main reason being the common prejudice in the particle physics community that BSM phenomena are likely to manifest themselves in the interactions between the two heaviest states of the SM, indeed $t$ and $h$, which are in fact intimately related to the hierarchy  problem of the SM, the main puzzle that Nature has forced upon us.

\begin{table} [!t]
  \centering
  \caption{The current experimental upper limits on top-Higgs FCNC decays at 95\% CL.}\label{table:top_fcnc_results}\vspace*{0.25cm} %
  \renewcommand{\arraystretch}{1.5}
  \begin{tabular}{c |c | c | c} \hline\hline
  \textbf{Detector}& Decay channel  & BR$(t \rightarrow u h)$ & BR$(t \rightarrow c h)$  \\ \hline
  \multirow{4}{*}{ATLAS, 13 TeV, 36.1 fb$^{-1}$ }&$h\to \gamma\gamma$~\cite{atlas13-1} & $2.4 \times 10^{-3}$ & $2.2 \times 10^{-3}$  \\
 &multilepton states~\cite{atlas13-2} & $1.9 \times 10^{-3}$ & $1.6 \times 10^{-3}$  \\  \
   &$h\to b\bar{b}$~\cite{atlas13} & $5.2 \times 10^{-3}$ & $4.2 \times 10^{-3}$  \\
  &$h\to \tau^{+}\tau^{-}$~\cite{atlas13} & $1.7 \times 10^{-3}$ & $1.9 \times 10^{-3}$  \\ \hline
   CMS, 13 Tev,  35.9 fb$^{-1}$& $h\to b\bar{b}$~\cite{cms13}& $4.7 \times 10^{-3}$ & $4.7 \times 10^{-3}$   \\ \hline
  \end{tabular}
\end{table}

In our present paper, we perform an updated study of top-Higgs FCNC interactions at the HE-LHC and FCC-hh, by considering both  single top quark production in  association with the  Higgs boson (plus a jet)  and top quark pair production followed by a Higgs decay of one (anti)top state. A previous study done in Ref.~\cite{Wu:2014dba} has investigated the top-Higgs FCNC interactions through $pp\to thj$ with
the subsequent decays $t\to b\ell^{+}\nu$ and $h\to \gamma\gamma$
at the HL-LHC. Here, we intend to revisit that analysis in the context of  the aforementioned higher energy and luminosity hadron machines.

Furthermore, past literature also  included the study of single top and Higgs boson associated production via the process $pp\to th$, affording one with an  improved sensitivity to especially the $tuh$ coupling (and somewhat less so to the $tch$ one) \cite{cms13}.
Specifically, the authors of Ref.~\cite{Liu:2016gsi}  investigated the top-Higgs FCNC interactions through the $pp\to t(\to b\ell^{+}\nu)h(\to \gamma\gamma)$ process at the HL-LHC. However, one realises that the final numbers of events for these signals at the 14 TeV LHC are too small
against the overwhelming SM background rate,
even considering the high luminosity option of 3 ab$^{-1}$, also because the signals suffer from a small BR (0.23\%) for the  $h\to \gamma\gamma$ channel. Yet, this is possibly the cleanest probe of  the SM-like Higgs boson, so it ought to be nonetheless explored.
In contrast, at both the HE-LHC and FCC-hh, the same production cross sections for signal (and SM background)  can be enhanced significantly due to the higher energies available therein, so that one can find it a more favourable environment than the 13 and 14 TeV LHC  to study the top-Higgs FCNC couplings via the  $h\to \gamma\gamma$ decay channel, at the same time  benefiting  a  larger luminosity.

In short, here, by studying the $pp\to thj$ process (i.e., with an explicit light jet in the final state\footnote{So that a direct comparison with existing experimental results can be made in the case of the $h\to b\bar{b}$ analysis of Ref.~\cite{atlas13}.}) inclusively, we aim at treating on the same footing both single and double top production.
This paper is arranged as follows. In Sec.~II, we give a brief introduction to the top-Higgs FCNC couplings and perform a complete calculation of $pp\to thj$ by considering such interactions at tree level. In Sec.~III, we discuss the observability (against the SM  background) of such top-Higgs FCNC couplings through the process $pp \to thj$ with the top  producing leptonic decay modes accompanied by $h\to \gamma\gamma$ at the HE-LHC and FCC-hh. Finally, conclusions and outlook are presented in Sec.~IV.

\section{Top-Higgs FCNC interactions and production processes}
\subsection{Top-Higgs FCNC couplings}
Although the anomalous FCNC couplings between the top quark and Higgs boson may arise from different sources, an effective field theory approach can describe the effects of NP beyond the SM in a model-independent way~\cite{AguilarSaavedra:2004wm}.  The most general Lagrangian for the top-Higgs FCNC interactions is written as
\begin{equation}
{\cal L}= \kappa_{tuH}\bar{t}Hu+\kappa_{tcH}\bar{t}Hc+h.c.,
\label{tqh}
\end{equation}
where $\kappa_{tuH}$ and $\kappa_{tcH}$ represent the strength of top-Higgs FCNC interactions. In this study we take them as real and symmetric, i.e., $\kappa_{tqH}=\kappa^{\dagger}_{tqH}=\kappa_{qtH}=\kappa^{\dagger}_{qtH}$ ($q=u,c$), since we here do not intend to consider  CP-violating effects.

The decay width of the dominant top quark decay mode $t\rightarrow Wb$ could be found in Ref.~\cite{Li:1990qf}. Neglecting the light quark masses and assuming the dominant top decay width $t \to Wb$, the Next-to-Leading Order (NLO) BR($t \to qh$) is given by~\cite{Greljo:2014dka,Liu:2018bxa}
\begin{equation}
{\rm BR}(t \to qh) = \frac{\kappa^{2}_{tqH}}{\sqrt{2} m^2_t G_F}\frac{(1-x^2_h)^2}{(1-x^2_W)^2 (1+2x^2_W)}\lambda_{\rm QCD} \simeq 0.58\kappa_{tqh}^{2},
\label{br}
\end{equation}
with the Fermi constant $G_F$ and $x_i=m_i/m_t~(i=W,\ h)$.  Here the factor $\lambda_{\rm QCD}$ is the NLO QCD correction to BR$(t \to qh)$ and equals about 1.1~\cite{Zhang:2008yn,Drobnak:2010wh,Zhang:2013xya}. In our work, we require $\kappa_{tqh}\leq 0.04$
to satisfy the direct constraint from the ATLAS result mentioned in the previous section.

\subsection{Production processes}
\begin{figure}[!t]
\begin{center}
\centerline{\epsfxsize=14cm \epsffile{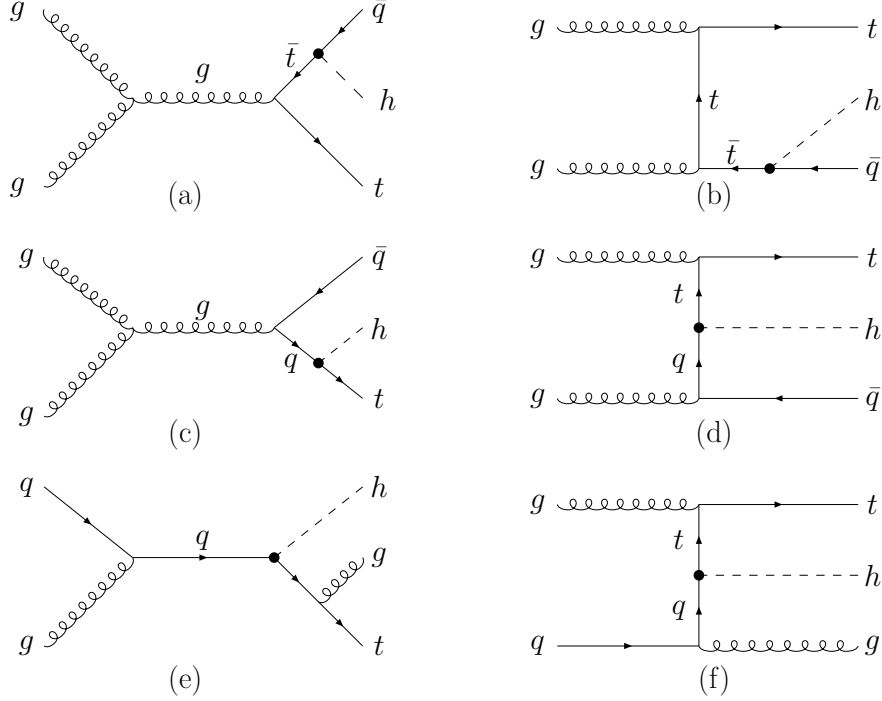}}
\vspace{-9cm}
\caption{Representative Feynman diagrams for: the $gg$ fusion induced top pair production $gg\to t\bar t$ and $\bar t\to \bar qh$ decay (a-b),  the $gg$ fusion induced top-Higgs  associated production $gg\to th\bar q$  (c-d) and the $qg$ fusion induced top-Higgs associated production $qg\to thg$ (e-f). Here $q=u,c$.}
\label{fey}
\end{center}
\end{figure}
At the LHC, the cross section for $pp\to thj$ involving top-Higgs FCNC couplings would be coming from two  subprocesses: (i) top pair production followed by one FCNC top decay, $pp\to t\bar{t}\to thj$,  shown in Fig.~1(a-b) (henceforth referred to as `top FCNC decay'); (ii) single top-Higgs associated production in presence of a jet, $pp\to thj$, as shown in Fig.~1(c-f), which includes a $gg$  (henceforth referred to as `tH associated production') and a $qg$  (henceforth referred to as `qg fusion') induced subchannels, respectively  yielding a(n) (anti)quark or gluon in the final state. The contribution of other subprocesses, such as $q\bar{q}$ fusion channels, is smaller than the above ones due to the suppression from colour factors  and Parton Distribution Functions (PDFs) and thus is not shown in the Feynman diagrams, but all the contributions are included in our calculations. Obviously, the conjugated processes can also occur at  tree level and are accounted for.

For the simulations of the
HE-LHC and FCC-hh dynamics, we first use
the \texttt{FeynRules} package~\cite{feynrules} to extract the Feynman rules
from the effective Lagrangian and to generate the Universal
FeynRules Output (UFO) files and calculate the LO cross sections of $pp\to thj$ by using \texttt{MadGraph5-aMC$@$NLO}~\cite{mg5} with \texttt{NNPDF23L01} PDFs~\cite{Ball:2014uwa}, considering the renormalisation and factorisation scales to be $\mu_R=\mu_F=\mu_0/2=(m_t + m_h)/2$. In our numerical calculations,
the SM input parameters are taken as~\cite{pdg}:
\begin{align}
m_h&=125.1{\rm ~GeV}, \quad m_t=172.9{\rm ~GeV}, \quad m_W=80.379{\rm ~GeV},\\ \nonumber
m_Z&=91.1876{\rm ~GeV}, \quad \alpha_{s}(m_Z)=0.1185, \quad G_F=1.166370\times 10^{-5}\ {\rm GeV^{-2}}.
\end{align}

\begin{figure}[htb]
\begin{center}
\vspace{-0.5cm}
\centerline{\epsfxsize=8cm\epsffile{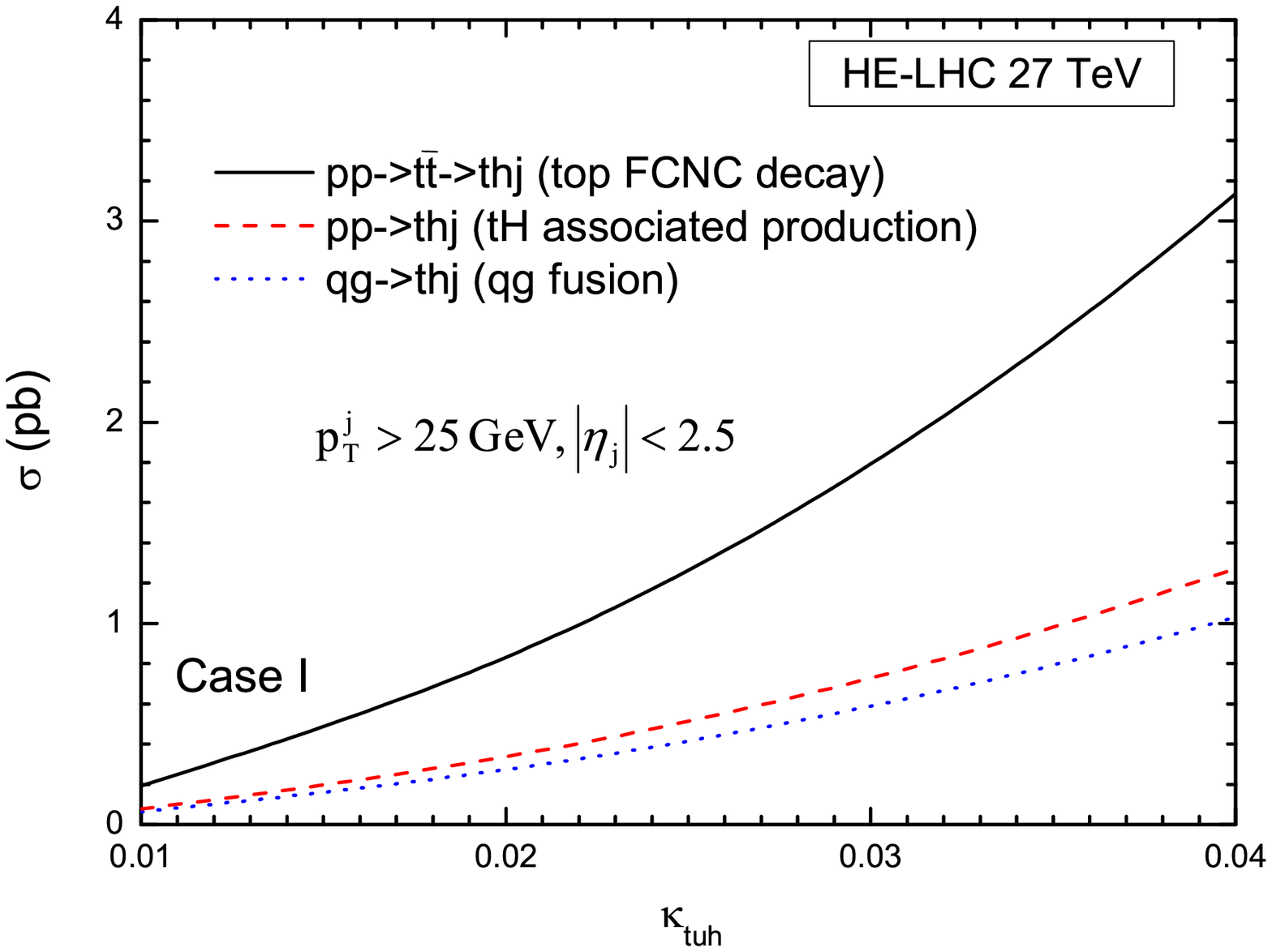}
\hspace{-0.5cm}\epsfxsize=8cm\epsffile{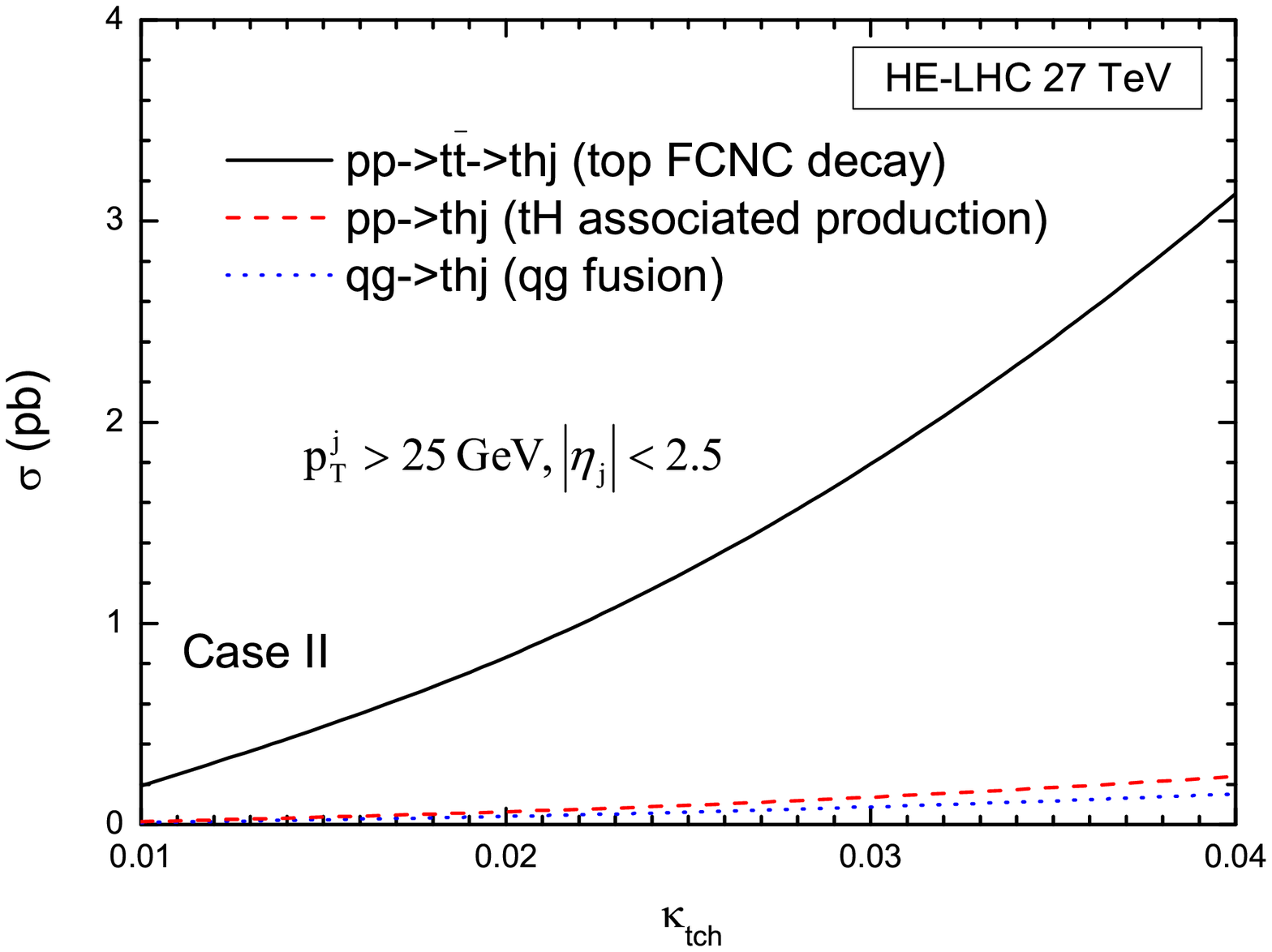}}
\centerline{\epsfxsize=8cm\epsffile{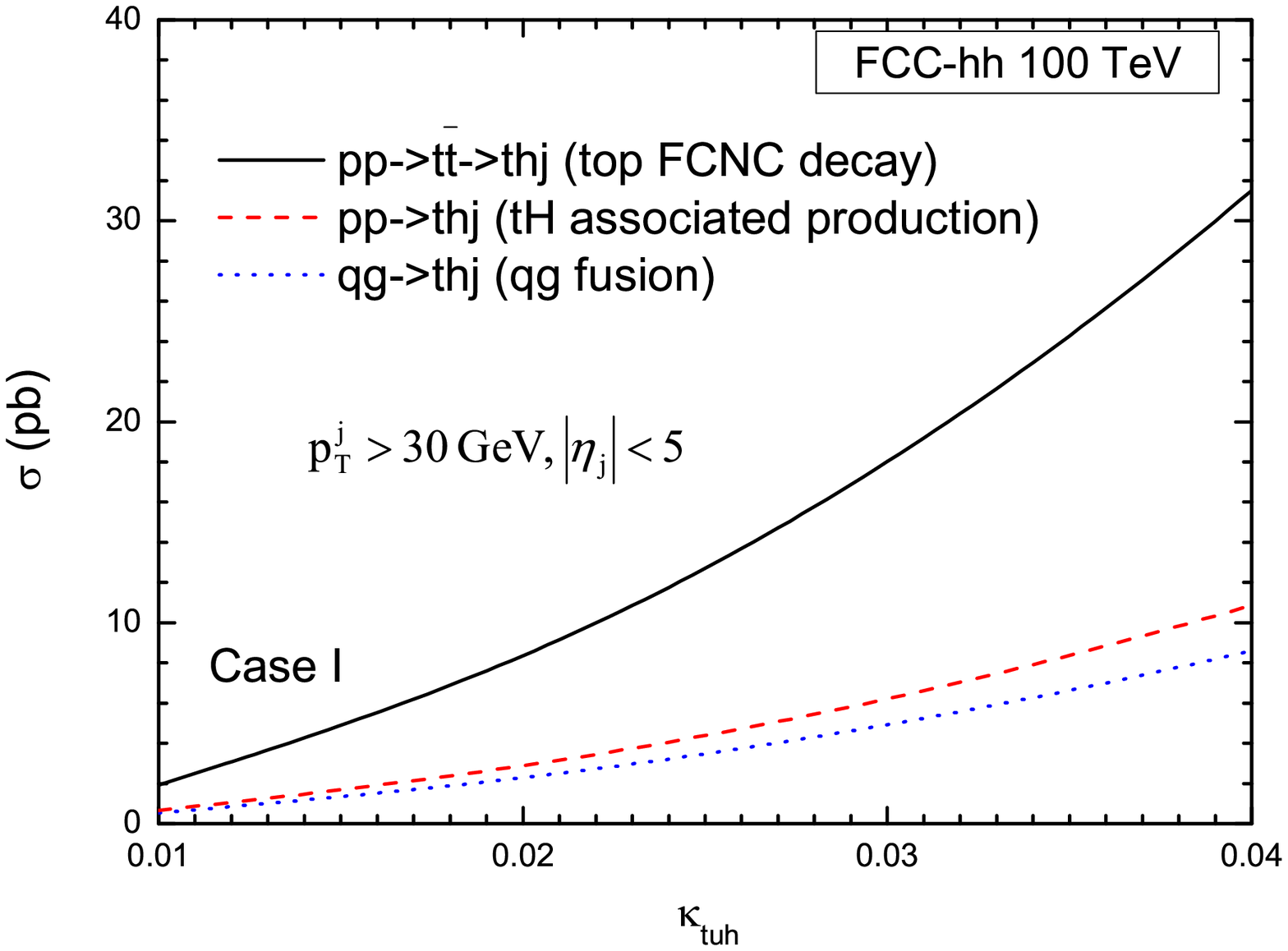}
\hspace{-0.5cm}\epsfxsize=8cm\epsffile{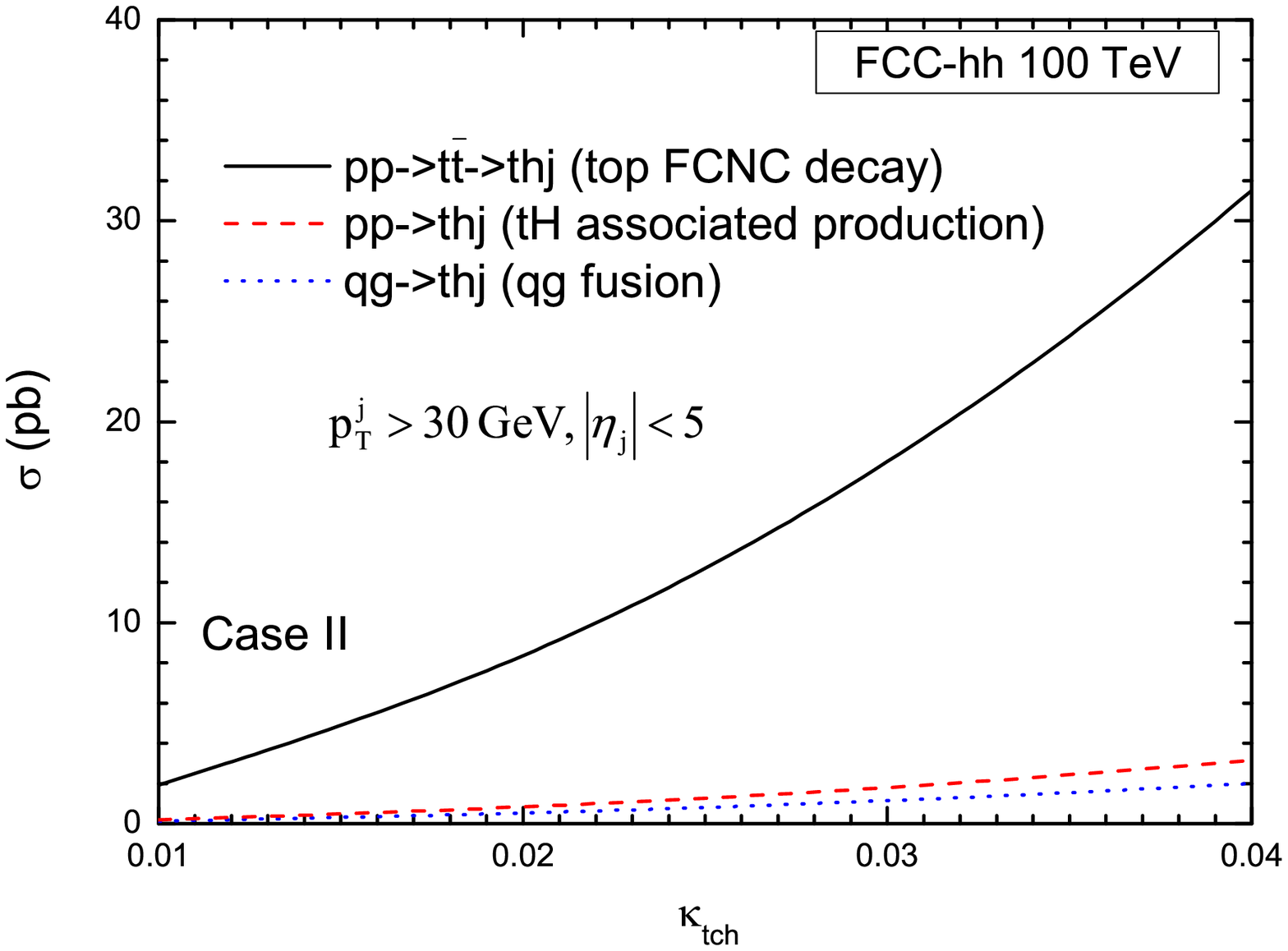}}
\caption{The dependence of the cross section $\sigma$ of the three $pp\to thj$ subprocesses of Fig.~1 on the top-Higgs FCNC couplings $\kappa_{tqh}$  at the HE-LHC (top) and FCC-hh (bottom) for Case I (left) and Case II (right) identified in the text. Notice that the charge conjugated  processes are also included in the calculation.}
\label{cross}
\end{center}
\end{figure}
In Fig.~\ref{cross}, we show the dependence of the cross sections for the  three $thj$ subprocesses on the top-Higgs FCNC coupling parameter at the HE-LHC and FCC-hh for two scenarios, as follows: Case I is for  $\kappa_{tqh}=\kappa_{tuh}, \kappa_{tch}=0$ whereas Case II is for $\kappa_{tqh}=\kappa_{tch}, \kappa_{tuh}=0$. {{The cuts on the transverse momentum ($p_T^j$) and  pseudo-rapidity ($\eta_j$) of the extra jet are shown in the figures for both the HE-LHC and FCC-hh.}}
From Fig.~\ref{cross} one can see that, for a given coupling parameter $\kappa_{tqh}$, the production cross sections can be very significant at the higher c.m. energies of these two future machine. Besides, we also have the following observations.
\begin{enumerate}
\item
For both Case I and II, the dominant contribution to the $thj$ final state is from  (resonant) pair production, $pp\to t\bar{t}\to thj$. However, the other two  contributions from tH associated production and qg fusion cannot be neglected, especially for Case I. To be specific, in this scenario, when $\sqrt{s}=27~(100)$ TeV and $\kappa_{tuh}=0.04$, the cross section of the top FCNC decay process is about 3.5~(30.9) pb while the cross section of the tH associated production process is about 1.7~(10.7) pb with the one for qg fusion being 1.3~(8.4) pb.
\item
For the same values of $\kappa_{tuh}$ and $\kappa_{tch}$, the cross sections coming from the tH associated production and qg fusion processes  in Case I are much larger than those in Case II: this is because the $u$-quark has a larger PDF than that of the  $c$-quark. To be specific, in Case II, when $\sqrt{s}=27~(100)$ TeV and $\kappa_{tch}=0.04$, the cross section of the tH associated production process is only about 0.3~(3.1) pb while for qg fusion the rates are 0.19~(1.96).
\end{enumerate}

\section{Discovery potential}
\subsection{The signal-to-background analysis}
In this section, we present the numerical calculations  at the HE-LHC and FCC-hh of the processes
\beq\label{signal}
pp&\to &t\bar{t}\to t(\to W^{+}b\to \ell^{+}\nu b)h(\to \gamma\gamma)j,\\
pp&\to &t(\to W^{+}b\to \ell^{+}\nu b)h(\to \gamma\gamma)j,
\eeq
where $\ell= e, \mu$ and $j$ represents (a(n) (anti)quark or gluon) jet, interfaced to the subsequent parton shower by using the  MLM matching scheme~\cite{Alwall:2007fs,Alwall:2008qv}. {{The final state topology is thus characterised by two photons appearing as a narrow resonance centered around the SM-like Higgs boson mass, at least two jets with exactly one being tagged as $b$-jet, one charged lepton and missing transverse momentum from the undetected neutrino.}} The main sources of background events that include both a Higgs boson decaying into di-photons in association with other particles and non-resonant production of $\gamma\gamma$ pairs are accounted for here:
 \begin{itemize}
\item
$pp\to t\bar{t}h$,
\item
$pp\to thj$,
\item
$pp\to W^{\pm}jjh$,
\item
$pp\to t\bar{t}\gamma\gamma$,
\item
$pp\to tj\gamma\gamma$,
\item
$pp\to \gamma\gamma W^{\pm}jj$.
\end{itemize}

 The parton level events for the signal and the SM backgrounds are interfaced to
parton shower, fragmentation and hadronisation by using  \texttt{PYTHIA8.20}~\cite{pythia8}. Then, we have passed all generated events through  \texttt{Delphes3.4.2}~\cite{delphes3} for detector
simulation.
Finally, event analysis is performed by using \texttt{MadAnalysis5} \cite{ma5}. As far
as  jet reconstruction is concerned, the anti-$k_{t}$ algorithm~\cite{Cacciari:2011ma} with a jet radius of 0.4 is used.
For the HE-LHC and FCC-hh analysis, we have used the default HL-LHC and FCC-hh detector card configuration
implemented into the aforementioned detector emulator.

 The cross sections of the signal and  dominant backgrounds at LO are adjusted to NLO QCD through $K$-factors, i.e., $K=1.4$ for the $pp\to t\bar{t}\to thj$ process~\cite{Mangano:2016jyj}, $K=1.5$ for the tH associated production process~\cite{Wang:2012gp,zhangcen} and $K=1.3$ for the $pp\to t\bar{t}h$ process~\cite{Mangano:2016jyj,Cepeda:2019klc,Yu:2014cka}. For the sake of simplicity, we have rescaled the other SM background processes by a $K$-factor
of 1.5. This approximation does not have a significant impact on our derived sensitivities  and can be fully addressed in a future analysis.

In order to identify objects, we impose the
following basic cuts to select the events~\cite{Mandrik:2018yhe}:
 \be
\mathrm{HE-LHC:} &
&p_{T}^{\ell/j/b}>&~25~\gev,
  &p_{T}^{\gamma}>&~20~\gev,
  &|\eta_{i}|<&~2.5,
  &\Delta R_{ij}>&~0.4 ~~(i,j = \ell, b, j, \gamma),\\
\mathrm{FCC-hh:}&
&p_{T}^{\ell/\gamma}>&~25~\gev,
  &p_{T}^{j/b}>&~30~\gev,
  &|\eta_{i}|<&~3,
  &\Delta R_{ij}>&~0.4 ~~(i,j = \ell, b, j, \gamma),\\
  \ee
where $\Delta R$ is the angular distance between any two objects.
\begin{figure*}[htb]
\begin{center}
\centerline{\epsfxsize=9.5cm\epsffile{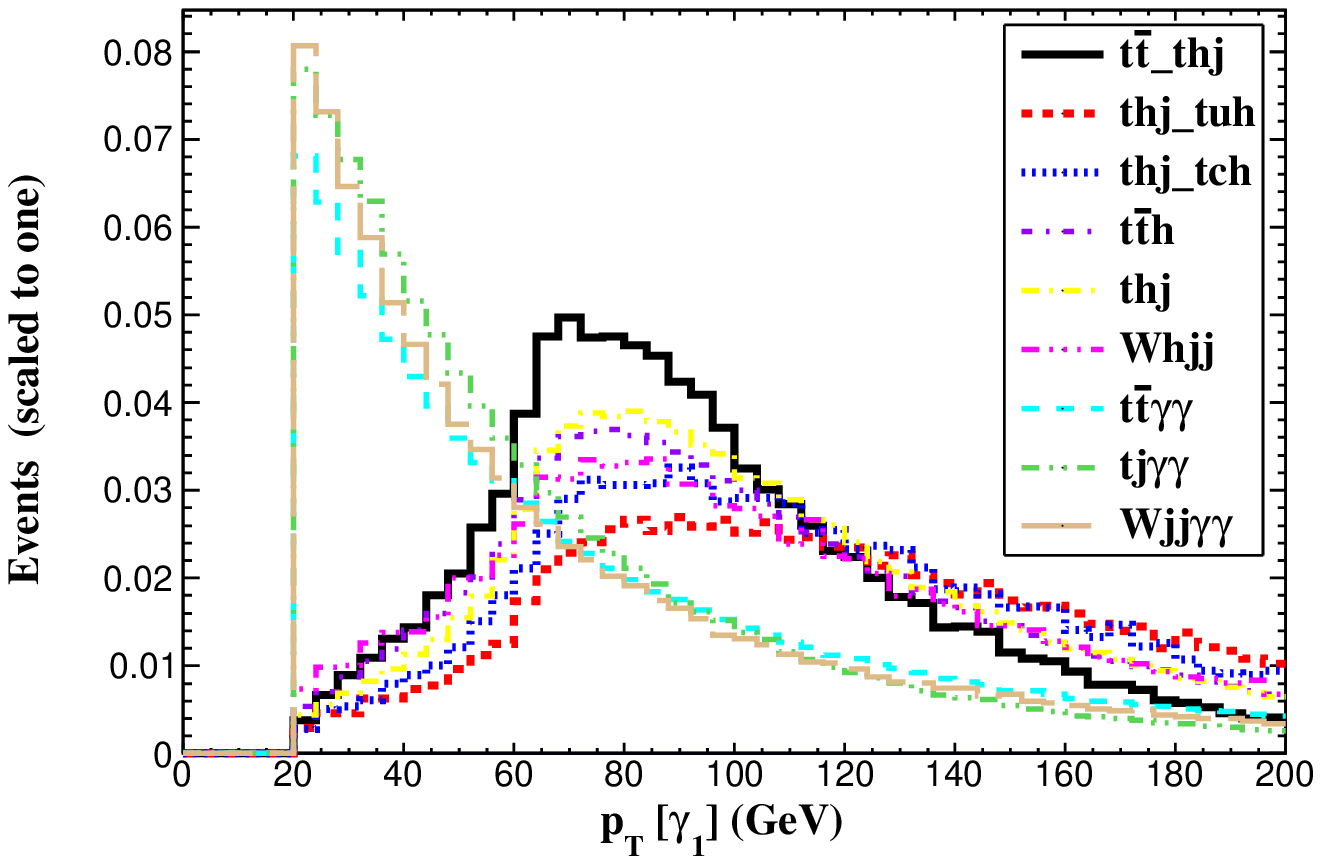}
\hspace{-2.0cm}\epsfxsize=9.5cm\epsffile{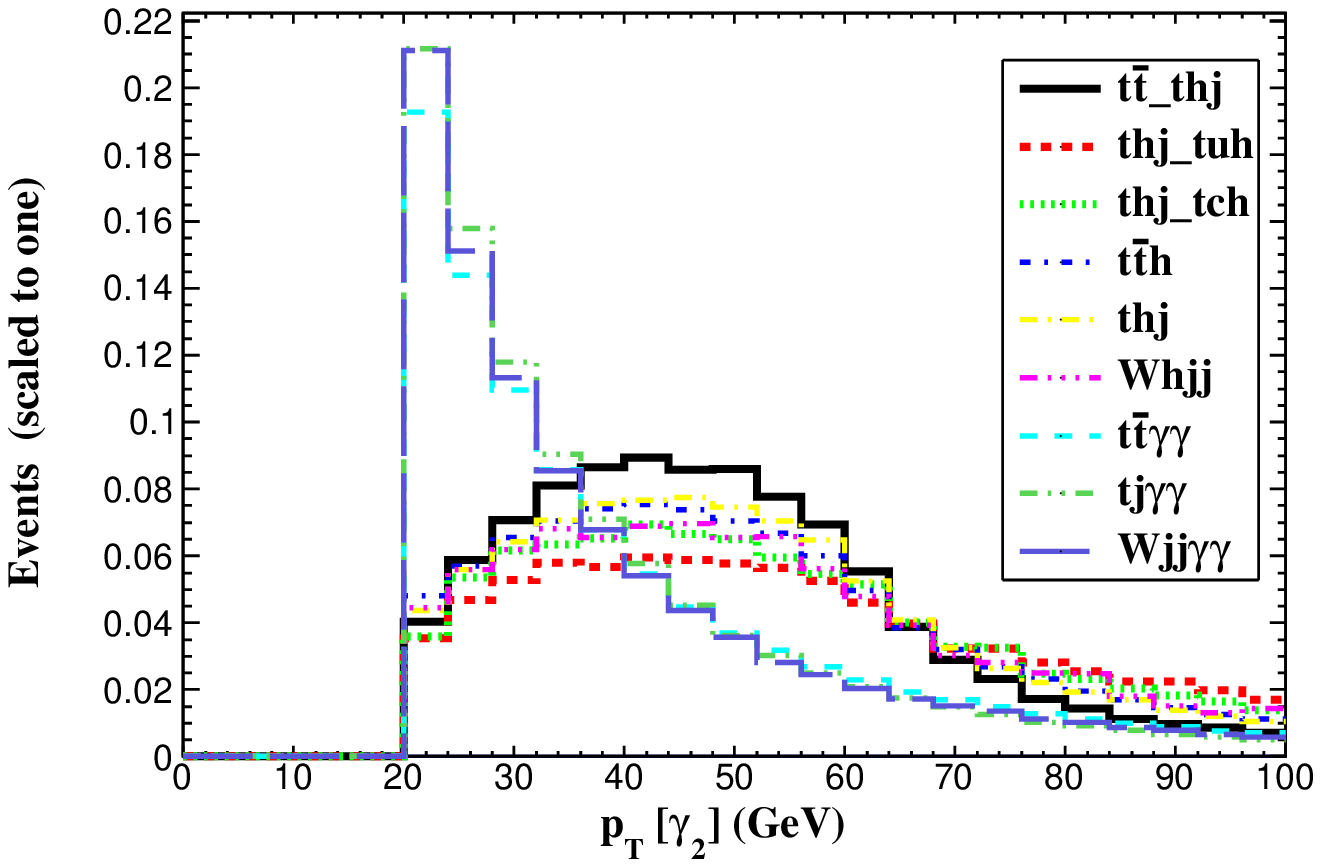}}
\centerline{\epsfxsize=9.5cm\epsffile{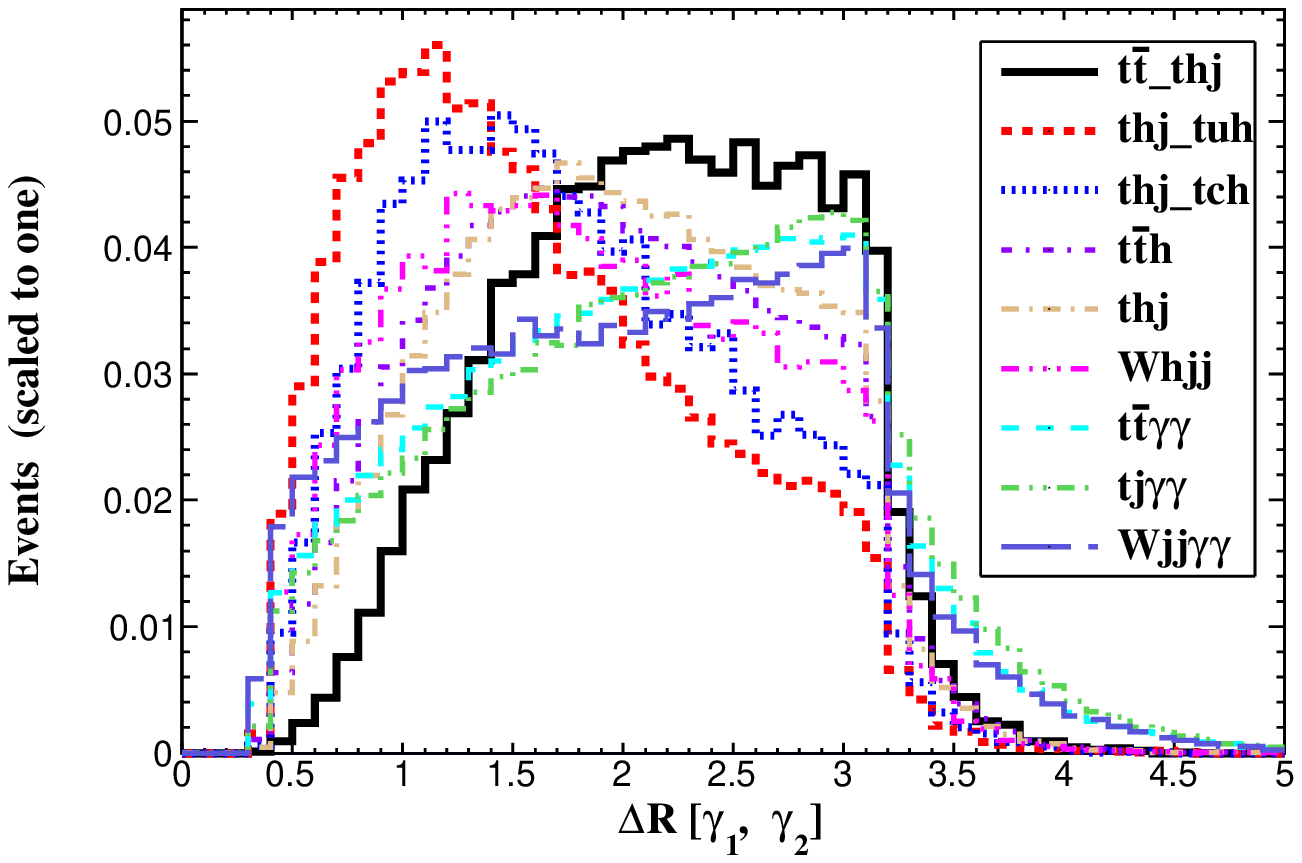}
\hspace{-2.0cm}\epsfxsize=9.5cm\epsffile{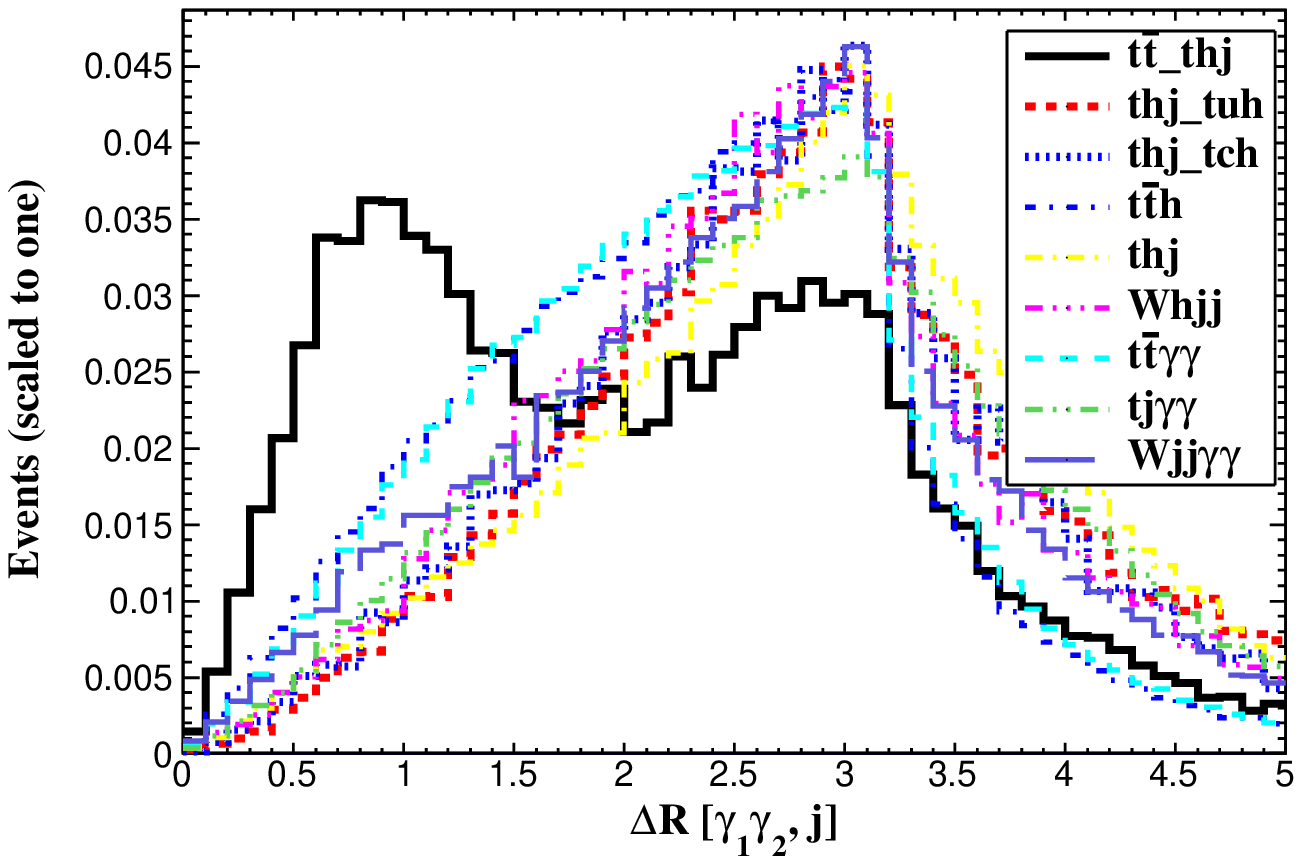}}
\centerline{\epsfxsize=9.5cm\epsffile{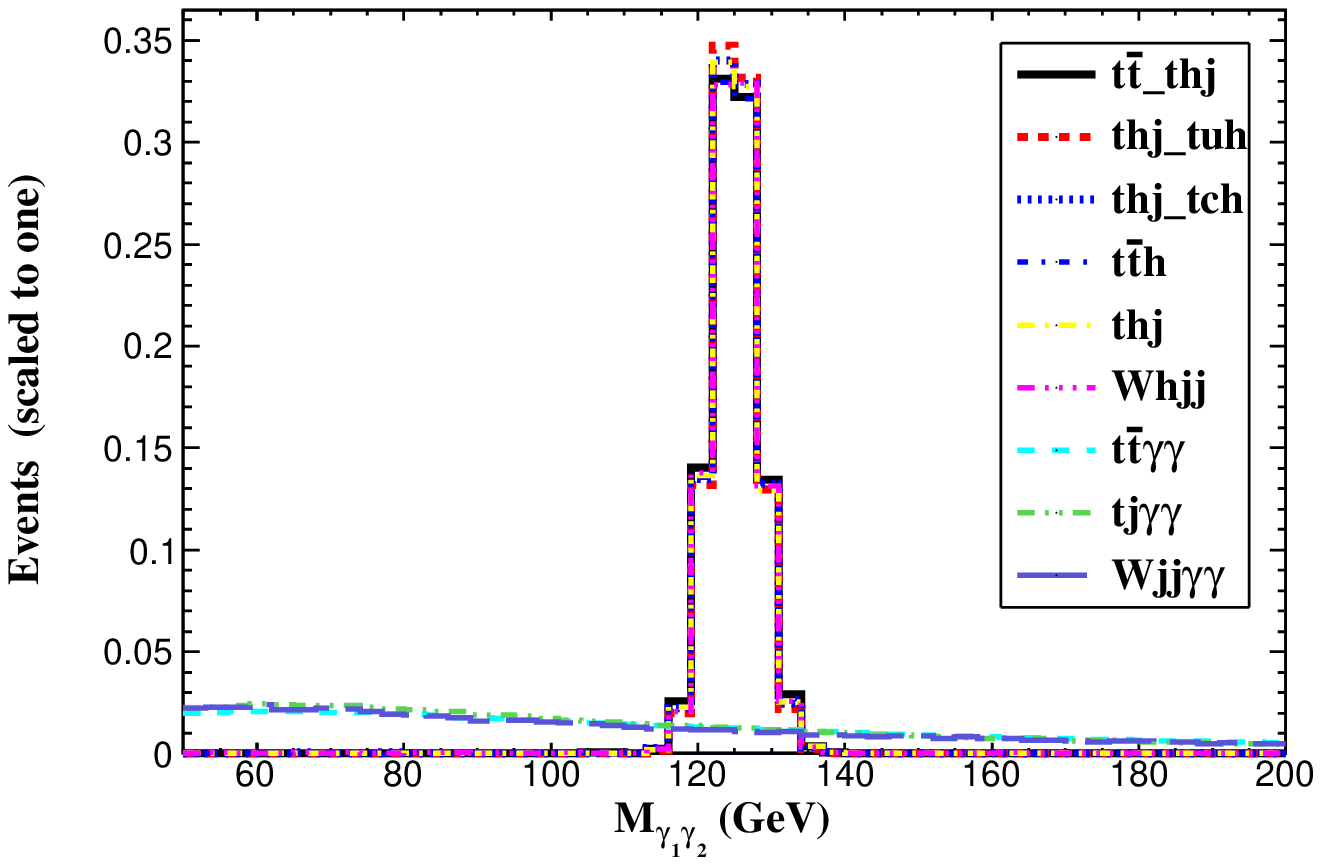}
\hspace{-2.0cm}\epsfxsize=9.5cm\epsffile{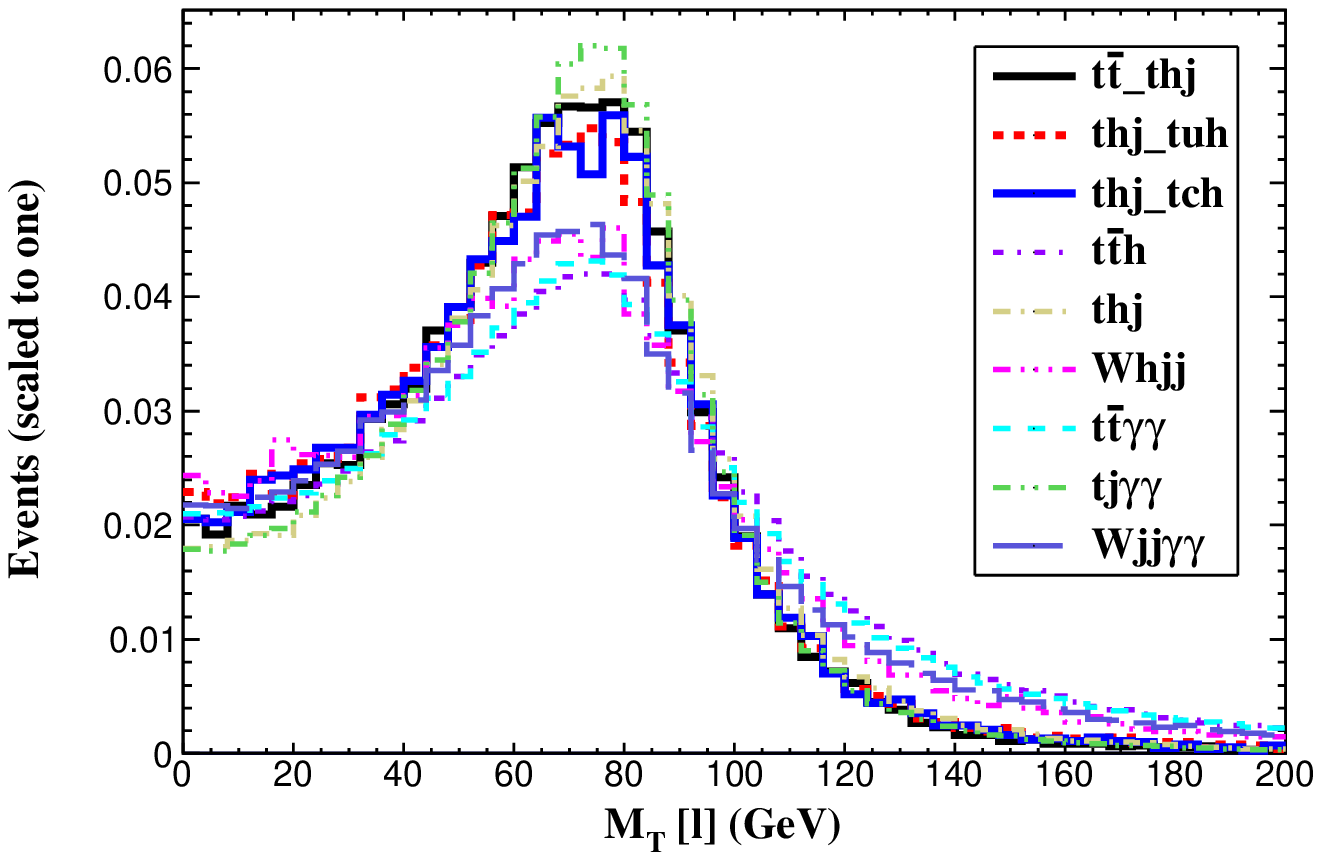}}
\centerline{\epsfxsize=9.5cm\epsffile{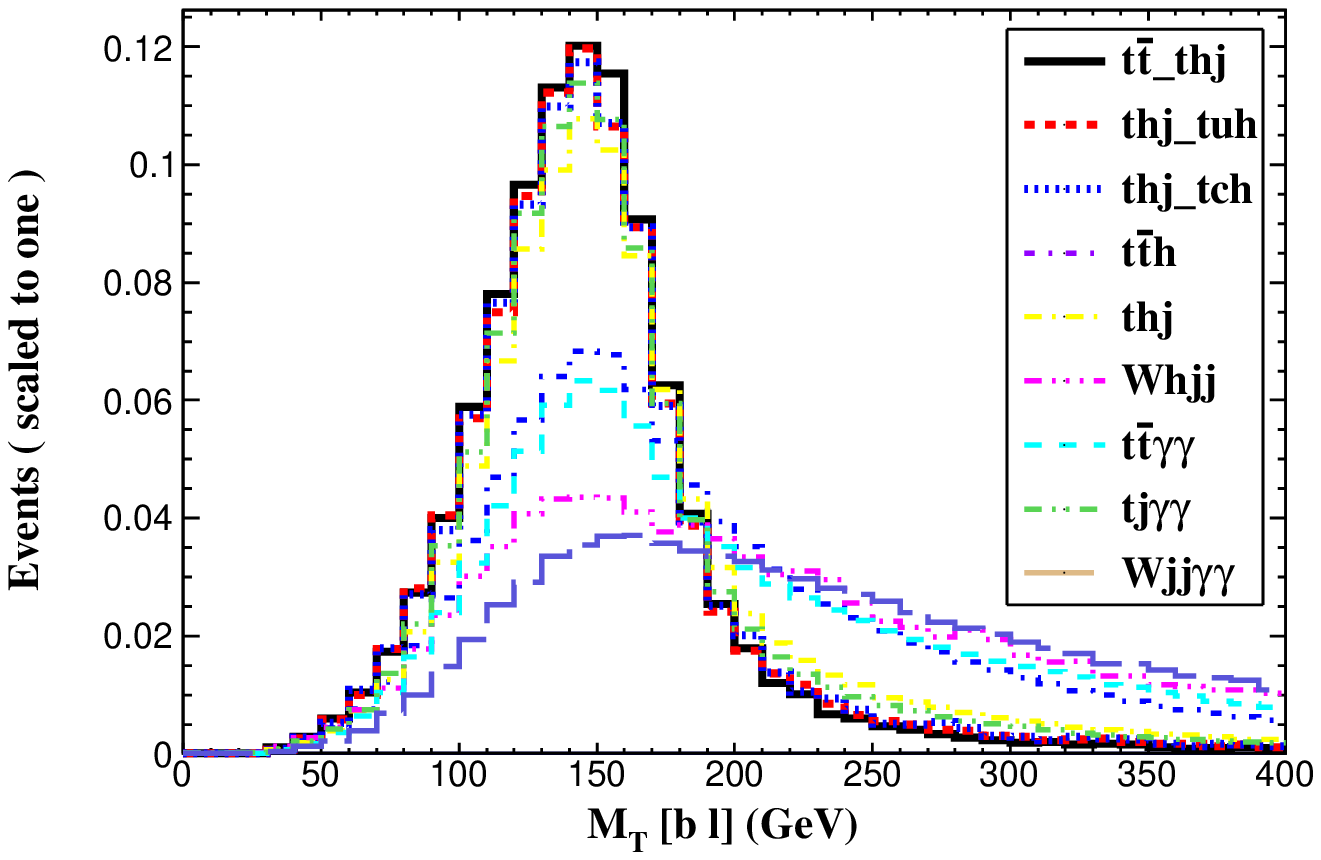}
\hspace{-2.0cm}\epsfxsize=9.5cm\epsffile{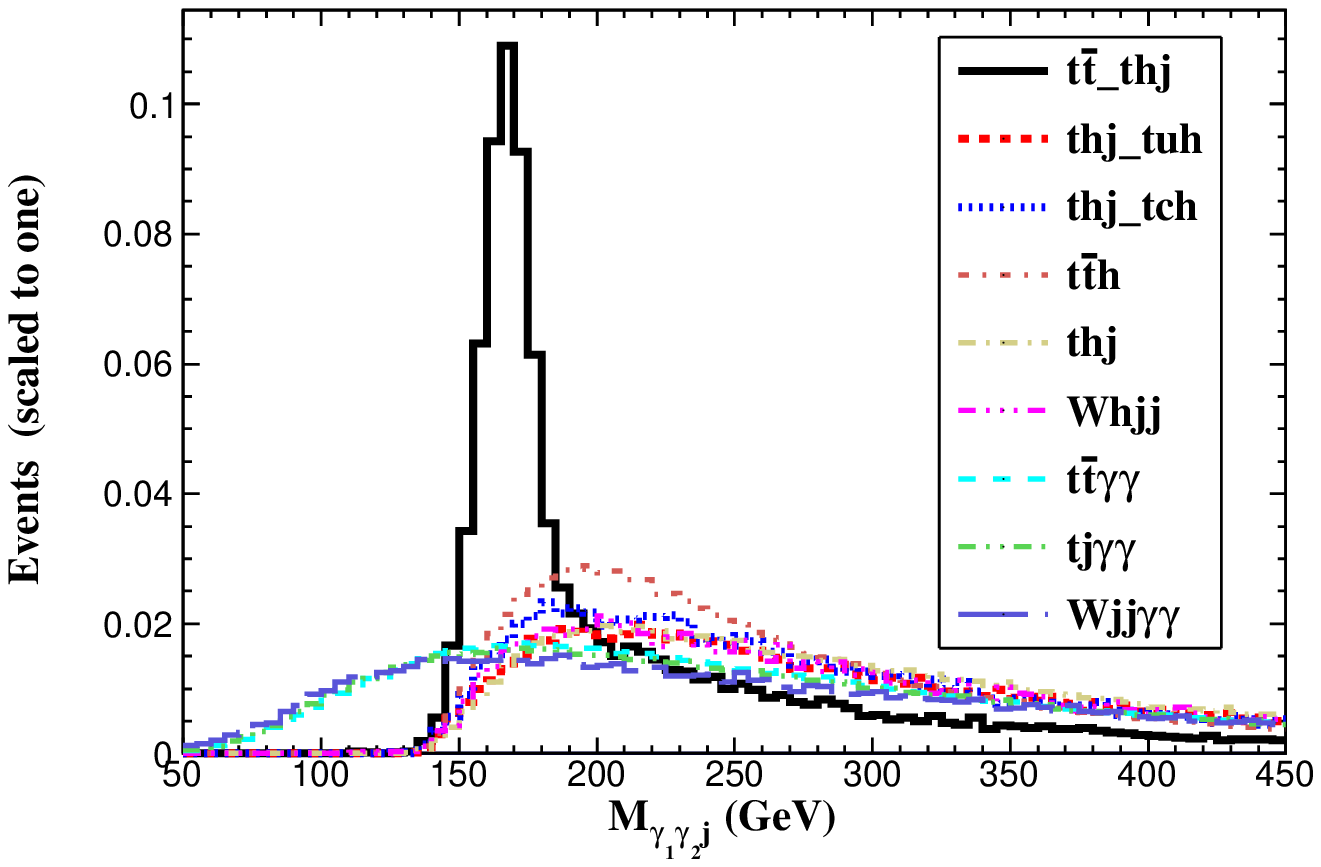}}
\caption{Normalised distributions for the signals and SM backgrounds at LO for the HE-LHC at 27 TeV.}
\label{distribution}
\end{center}
\end{figure*}

In order to choose appropriate kinematic cuts, in Fig.~\ref{distribution}\footnote[2]{Hereafter, in figures and tables, by using `thj\_tuh ($pp\to thj$ via $tuh$)' and `thj\_tch($pp\to thj$ via $tch$)', we will intend the contribution to the signal due to tH associated production plus $qg$ fusion when only including the $tuh$ or $tch$ coupling on its own, respectively.}, we plot some differential distributions   for  signals and SM backgrounds at the HE-LHC at 27 TeV, such as the (ordered) transverse momentum distributions of the two photons, $p_{T}^{\gamma_{1,2}}$, the separation, $\Delta R_{\gamma_{1},\gamma_{2}}$ and $\Delta R_{\gamma_{1}\gamma_{2},j}$, and invariant mass, $M_{\gamma\gamma}$, distributions of the two photons, the transverse mass distribution  for the $\ell \slashed E_T$, $M_{T}(l)$, and $b\ell \slashed E_T$, $M_{T}( bl)$  systems, and the invariant mass, $M_{\gamma\gamma j}$.
Based on these distributions, we impose a further set of cuts.
 \begin{itemize}
\item
Cut 1: Exactly one isolated lepton (electron  or muon), at least two jets and one of which must be $b$-tagged.
\item
Cut 2: At least two photons with $p_{T}^{\gamma_1}>60 \rm ~GeV$, $p^{\gamma_2}_{T}>30 \rm ~GeV$, since the two photons in the signal and  resonant SM backgrounds come from the Higgs boson they have a harder $p_T$ spectrum than those in the non-resonant SM backgrounds.
\item
Cut 3: The distance between two photons lies in $1.8 < \Delta R_{\gamma_{1},\gamma_{2}} <3.5$, the distance between the di-photon system and the extra light jet lies in $\Delta R_{\gamma\gamma,j} <1.8$.
\item
 Cut 4: The invariant mass of the di-photon system, $M_{\gamma\gamma}$, is peaked in both the signals and resonant backgrounds, thus we require $M_{\gamma\gamma}$ to be in the range $|M_{\gamma\gamma} - m_h| < 2$ GeV.
\item
 Cut 5: The transverse mass $M_T(\ell)$ and $M_T(b\ell)$ cuts are $M_T(\ell) > 30 \rm~GeV$ and
$100 ~{\rm GeV} < M_T(b\ell) < 180 \rm~GeV$.
\item
Cut 6: The invariant mass $M_{\gamma\gamma j}$ cut is $160 ~{\rm GeV} < M_{\gamma\gamma j} < 190 \rm~GeV$.
\end{itemize}

\begin{table}[htb]
\begin{center}
\caption{The cut flow of the cross sections (in ab) for the signals and SM backgrounds at the HE-LHC where the anomalous coupling parameters are taken as $\kappa_{tuh}=0.04$ or $\kappa_{tch}=0.04$ in the signal, while fixing the other to zero. }\label{cutflow-1}
\vspace{0.2cm}
\begin{tabular}{c|c|c|c|c|c|c|c|c}
\hline \hline
\multirow{2}{*}{Cuts}& \multicolumn{2}{c|}{Signal}&\multicolumn{6}{c}{Backgrounds} \\ \cline{2-9}
& $t\bar{t}\to thu~(thc)$ & $thu~(thc)$& $t\bar{t}h$ &$thj$ & $W^{\pm}jjh$ & $t\bar{t}\gamma\gamma$  & $tj\gamma\gamma$& $\gamma\gamma W^{\pm}jj$\\   \cline{1-9} \hline
Basic cuts &366~(358) &114~(37) &269 &35 &20&4064 &4880 &5985\\
Cut 1 &277~(268)&96~(29)&119&25&16&1819&3660&4988\\
Cut 2 &164~(157)&68~(19)&68&15&8&348&610&855\\
Cut 3 &40~(38)&3.6~(1.3)&10&1.3&0.7&56&74&109\\
Cut 4 &19~(18)&1.7~(0.6)&5&0.6&0.3&1.4&2.1&2.6\\
Cut 5 &13~(13)&1.3~(0.4)&2.2&0.4&0.1&0.6&1.2&0.3\\
Cut 6 &8.9~(7.7)&0.6~(0.2)&0.95&0.16&0.04&0.18&0.42&0.19\\
\hline
\end{tabular} \end{center}\end{table}

\begin{table*}[htb]
\begin{center}
\caption{The cut flow of the cross sections (in ab) for the signals and SM backgrounds at the FCC-hh where the anomalous coupling parameters are taken as $\kappa_{tuh}=0.04$ or $\kappa_{tch}=0.04$ in the signal, while fixing the other to zero. }\label{cutflow-2}
\vspace{0.2cm}
\begin{tabular}{c|c|c|c|c|c|c|c|c}
\hline \hline
\multirow{2}{*}{Cuts}& \multicolumn{2}{c|}{Signal}&\multicolumn{6}{c}{Backgrounds} \\ \cline{2-9}
& $t\bar{t}\to thu~(thc)$ & $thu~(thc)$& $t\bar{t}h$ &$thj$ & $W^{\pm}jjh$ & $t\bar{t}\gamma\gamma$  & $tj\gamma\gamma$& $\gamma\gamma W^{\pm}jj$\\   \cline{1-9} \hline
Basic cuts &4053~(4364) &1187~(538) &7575 &592 &324&73800 &63000 &55158\\
Cut 1 &3262~(2876)&899~(379)&2461&376&252&22140&40320&42795\\
Cut 2 &1879~(1686)&612~(257)&1285&215&144&5535&8820&20446\\
Cut 3 &440~(367)&28~(14)&161&15&10&775&932&2330\\
Cut 4 &396~(327)&25~(12)&142&14&9&16&22&45\\
Cut 5 &282~(221)&18~(9)&59&9&2.6&6.3&13&9\\
Cut 6 &222~(173)&7~(4)&21.6&3&1&1.6&4.2&3.6\\
 \hline
\end{tabular} \end{center}\end{table*}

For the FCC-hh
analysis at 100 TeV, we use the same selection cuts for the signal and SM backgrounds because the distributions are very similar to the case of HE-LHC presented in Fig.~\ref{distribution}. In fact, the difference between the HE-LHC and FCC-hh mainly
comes from the different detector configurations.
The effects of the suitable cuts on the signal and SM  background processes are illustrated in Tab.~\ref{cutflow-1} and Tab.~\ref{cutflow-2} at the HE-LHC and FCC-hh, respectively.  {{Due to the different $b$-tagging  rates for $u$- and $c$-quarks, the signal efficiencies of the two top (anti)quark
decays differ after applying requirements on the $b$-tagged jet
multiplicity. Thus we give the events separately for $q = u, c$. One can see that, at the end of the cut flow, the largest SM  background is the $pp\to t\bar{t}h$ process, which is about 0.001 fb and 0.02 fb at the HE-LHC and FCC-hh, respectively. Besides, the $tj\gamma\gamma$ process can also generate significant contributions to the SM background due to the large production cross section. Finally,
notice that, after the final cuts, the dominant signal contribution comes from the FCNC top (anti)quark decay process, so that one can safely ignore the contribution from the top-Higgs associated production channels. }}

\subsection{Sensitivities at the HE-LHC and FCC-hh}

To estimate the exclusion significance,
$Z_\text{excl}$,
we use the following expression~\cite{Cowan:2010js,Kumar:2015tna,Kling:2018xud}:
\be
    Z_\text{excl} =\sqrt{2\left[s-b\ln\left(\frac{b+s+x}{2b}\right)
  - \frac{1}{\delta^2 }\ln\left(\frac{b-s+x}{2b}\right)\right] -
  \left(b+s-x\right)\left(1+\frac{1}{\delta^2 b}\right)},
 \ee
with $x=\sqrt{(s+b)^2- 4 \delta^2 s b^2/(1+\delta^2 b)}$. Here, the values of $s$ and $b$ were obtained by multiplying the total signal and SM background cross
sections, respectively,  by the integrated luminosity. Furthermore, $\delta$ is the percentage systematic error on the SM
background estimate. In the limit of
$\delta \to 0$,  this expression  can be simplified as
\be
 Z_\text{excl} = \sqrt{2[s-b\ln(1+s/b)]}.
 \label{s95}
\ee
\noindent In this work we choose two cases: no systematics ($\delta=0$) and a systematic
uncertainty of $\delta=10\%$ for both the HE-LHC and  FCC-hh.
We define the regions with $Z_\text{excl} \leq 1.645$ as those that can be excluded at  95\% CL ($p=0.05$).
The limits on the FCNC coupling parameter $\kappa_{tqh}$ can be directly translated in terms of
constraints on BR$(t\to qh)$ by using eq.~(\ref{br}).

\begin{figure}[htb]
\begin{center}
\vspace{-0.5cm}
\centerline{\epsfxsize=8cm \epsffile{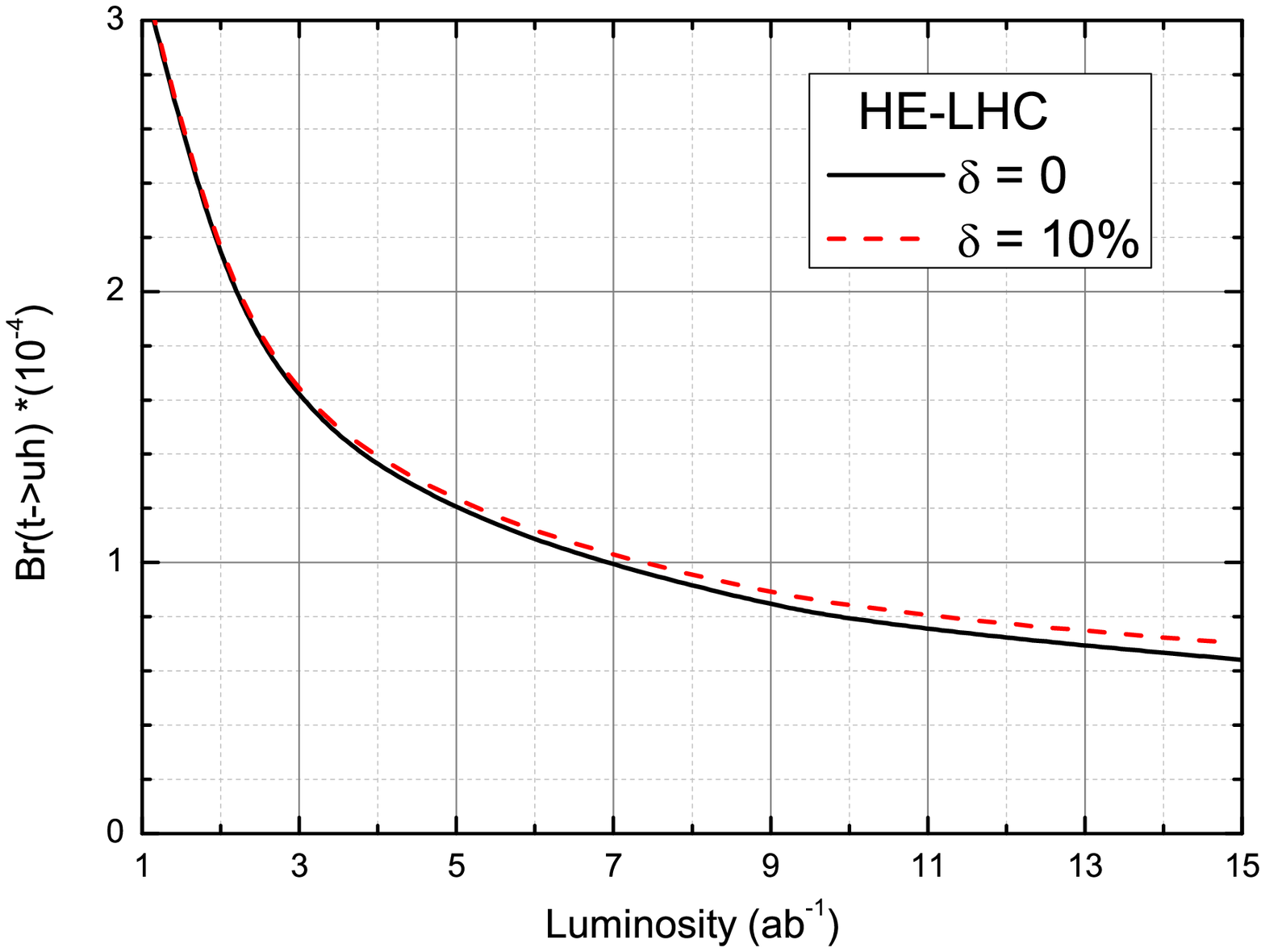}\epsfxsize=8cm \epsffile{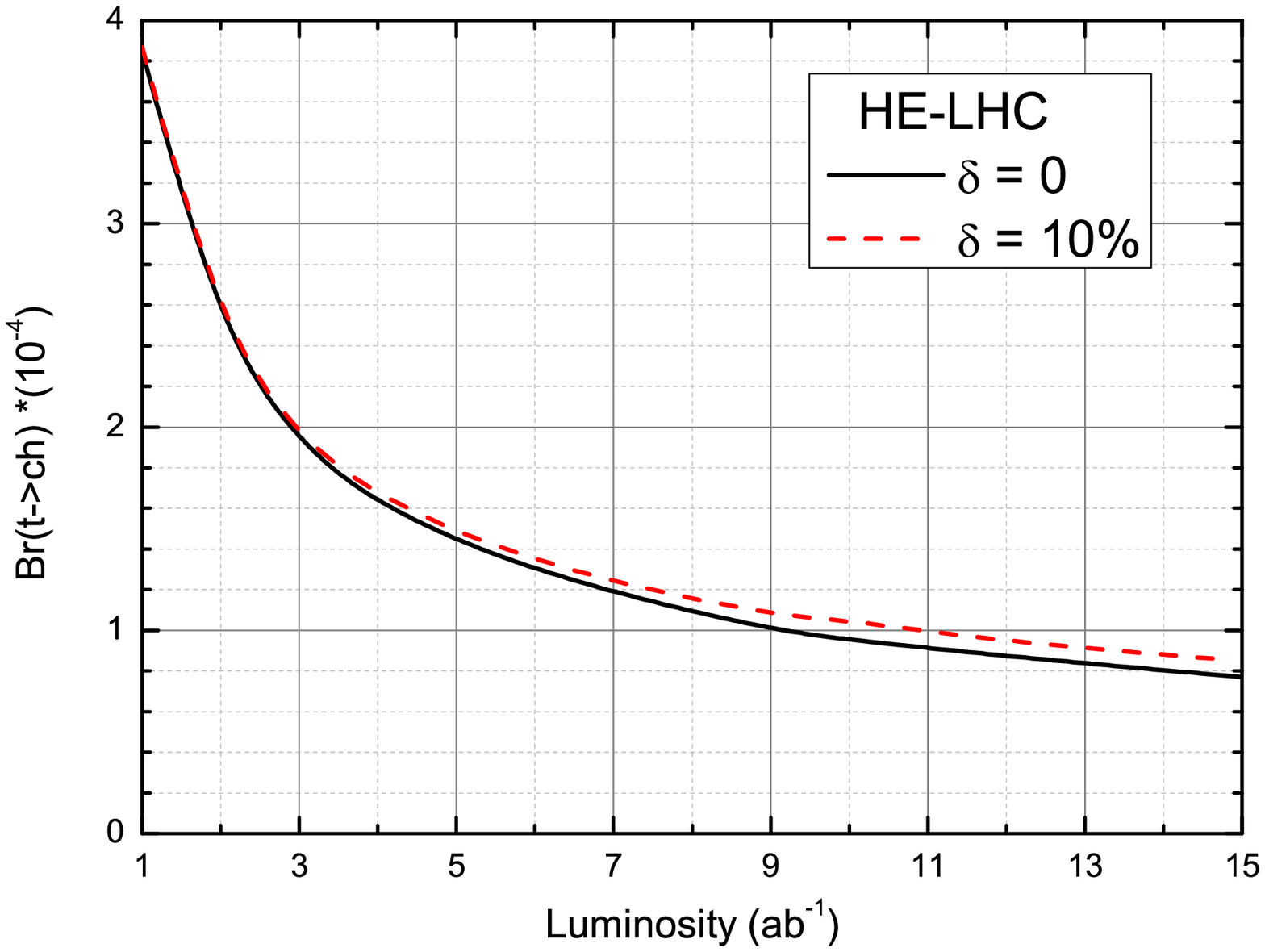}}
\caption{The exclusion limits at 95\% CL on BR$(t\to uh)$~(left) and BR$(t\to ch)$~(right) at the HE-LHC with two systematic error cases: $\delta=0$ and $\delta=10\%$.}
\label{ss95-HELHC}
\end{center}
\end{figure}

\begin{figure}[htb]
\begin{center}
\vspace{-0.5cm}
\centerline{\epsfxsize=8cm \epsffile{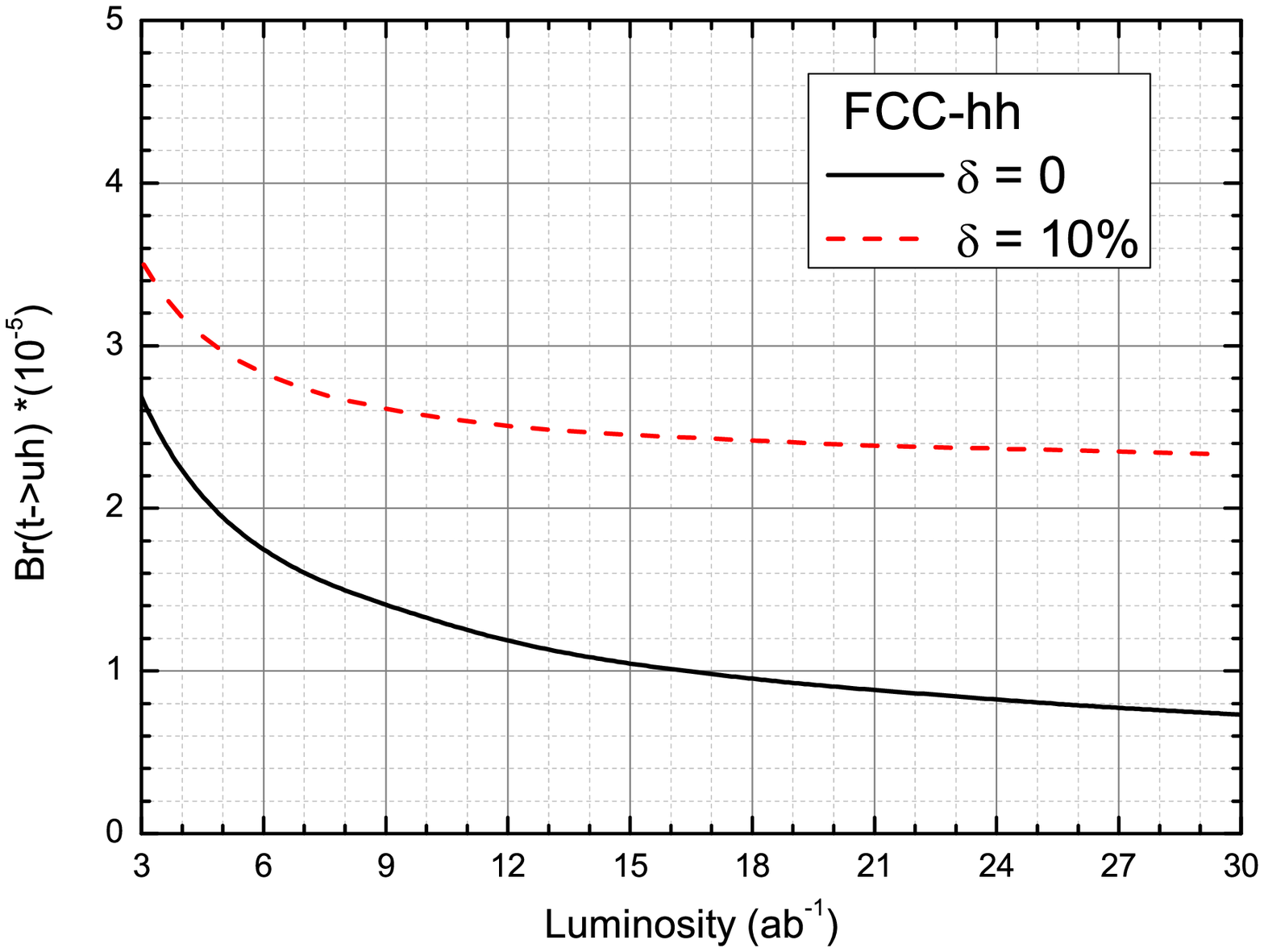}\epsfxsize=8cm \epsffile{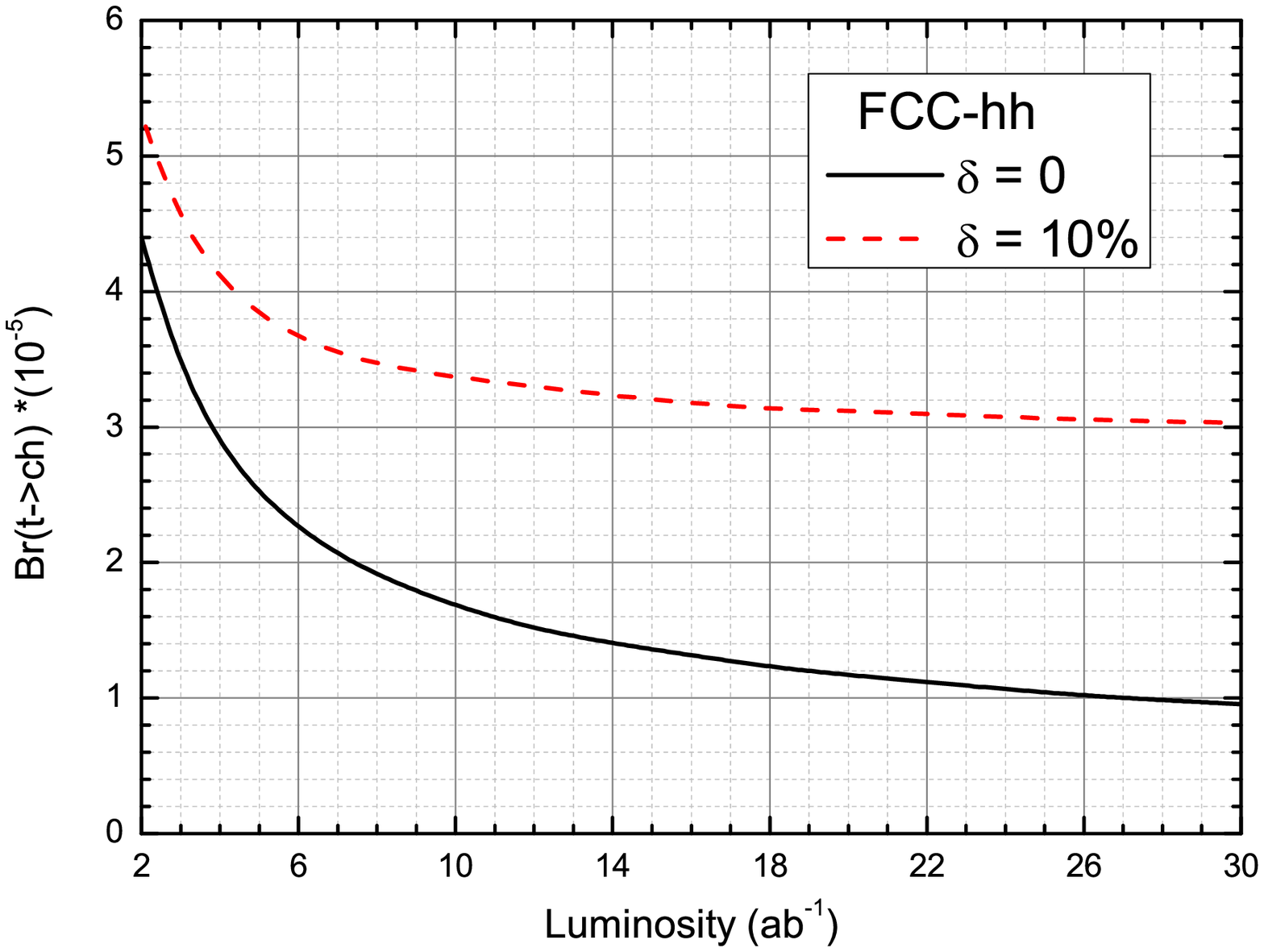}}
\caption{Same as Fig.~\ref{ss95-HELHC} but for the FCC-hh.}
\label{ss95-FCC}
\end{center}
\end{figure}

\begin{table*}[htbp]
\begin{center}
\caption{The upper limits on BR$(t\to qh)$ at 95\% CL obtained at the HE-LHC and FCC-hh. We consider systematic errors of 0\%
and 10\% on the SM background events only.  The 95\% CL upper limits obtained at the HL-LHC at 3 ab$^{-1}$ by the ATLAS Collaboration also have been shown for comparisons.\label{limit}}
\vspace{0.2cm}
\begin{tabular}{c| c| c| c |c |c}  \hline
\multirow{2}{*}{Branching fraction}& \multicolumn{2}{c|}{HE-LHC, 15 ab$^{-1}$}&\multicolumn{2}{c|}{FCC-hh, 30 ab$^{-1}$}  &\multirow{2}{*}{HL-LHC, 3 ab$^{-1}$} \\ \cline{2-5}
 & $\delta=0$  & $\delta=10\%$ &$\delta=0$  & $\delta=10\%$&\\   \cline{2-5} \hline
BR$(t \to u h)$  &   $6.4 \times 10^{-5}$  &   $7.0\times 10^{-5}$ &   $7.3 \times 10^{-6}$ &   $2.3 \times 10^{-5}$&     $2.4 \times 10^{-4}$, $h\to b\bar{b}$~\cite{atlas-14-3000} 	\\
BR$(t \to c h)$   &   $7.7 \times 10^{-5}$&   $8.5 \times 10^{-5}$ &   $9.6 \times 10^{-6}$ &   $3.0 \times 10^{-5}$  &   $1.5 \times 10^{-4}$, $h\to \gamma\gamma$~\cite{atlas1-14-3000}	\\
\hline
\end{tabular}
\end{center}
\end{table*}

In Figs.~\ref{ss95-HELHC}-\ref{ss95-FCC}, we plot the exclusion limits at 95\% CL in the plane of the integrated luminosity and the BR$(t\to qh)$'s at the HE-LHC and FCC-hh with the aforementioned two systematic error cases of  $\delta=0$ and $\delta=10\%$. One can see that,  our signals are rather robust against the systematic uncertainties
on the background determination,  {{ though they differ between the HE-LHC
(where limits change within a factor of $\approx1.1$)
and  FCC-hh (where limits change within a factor of $\approx3.1$)
due to the relatively different number of SM background events}}. The values for 95\% CL upper limits are summarised in
Tab.~\ref{limit}.
With a realistic 10\% systematic error, the sensitivities are slightly weaker than those without any systematic error, being of the order of $10^{-5}$ at the 95\% CL both at the HE-LHC and FCC-hh. For comparison, the recent 95\% upper limits on BR$(t\to qh)$ obtained at the HL-LHC with an integrated luminosity of 3 ab$^{-1}$ by the ATLAS Collaboration~\cite{atlas-14-3000} are also presented, which are obtained via the decay mode $t\to qh(\to b\bar{b})$.
Besides, the upgraded ATLAS experiment has also estimated top-Higgs FCNC couplings via the decays $t\to ch(\to \gamma\gamma)$ at the  HL-LHC and obtained an expected upper limit of BR$(t\to ch) < 1.5\times 10^{-4}$ at 95\% CL~\cite{atlas1-14-3000}. Altogether, the sensitivity to the BR of the $t\to qh$ are two order of magnitude better than the most recent direct limits reported by the ATLAS Collaboration at the $13 \, {\rm TeV}$ LHC.

Before closing, let us also review competing limits from other authors.
Very recently, the author of Ref.~\cite{Khanpour:2019qnw} has studied the top-Higgs  FCNC couplings in the triple-top signal at the HE-LHC and FCC-hh. The 95\% CL  upper limits on
BR$(t\to uh)$ (and BR$(t\to ch)$) were found, respectively,  as $7.01 \times 10^{-4}~(3.66 \times 10^{-4})$ at the
HE-LHC with 15 ab$^{-1}$ and as $2.49\times 10^{-5}~(5.85\times 10^{-5})$ at the FCC-hh with 10 ab$^{-1}$. Furthermore,
in the context of the 2-Higgs Doublet Model (2HDM), the authors of Ref.~\cite{Jain:2019ebq} have recently investigated the prospect for $t\to ch$ decay in top quark pair production via the $h\to WW^{\ast}\to \ell^{+}\ell^{-}+\slashed E_T^{miss}$ channel. For the HE-LHC and FCC-hh, the 95\% CL  upper limits on BR$(t\to ch)$ was found to be at the order of $10^{-4}$ with an integrated luminosity of 3 ab$^{-1}$ and such limits would be increased by an higher integrated luminosity. Finally, at the FCC-hh with an integrated luminosity of 10 ab$^{-1}$, Ref.~\cite{Papaefstathiou:2017xuv} has investigated the $t\to ch$ decay and the its BR can be constrained to $\mathcal{O}(10^{-5})$ either with or without considering $c$-jet tagging.

\section{Summary}
In this work, we have analysed the process $pp\to thj$ at the HE-LHC and FCC-hh by considering $thq$ FCNC couplings. We have performed a full Monte Carlo simulation for the
signals obtained
from three different subprocesses
via the top leptonic decay mode and $h\to \gamma\gamma$ against all relevant SM backgrounds. After a dedicated cut based selection, we have found that top pair production followed by one FCNC top decay is significantly more abundant than FCNC single top-Higgs associated production in presence of a jet.
The obtained exclusion limits on the $tqh$ coupling strengths and the ensuing  BRs  have been summarised and compared in detail to results in  literature, namely, the most recent  LHC experimental limits and the (projected) HL-LHC ones as well. {{Our results show that 95\% CL limits on the  BR$(t\to qh)$, with $q=u~(c)$,  have been found to be $6.4~(7.7) \times 10^{-5}$ at the HE-LHC with an integrated luminosity of 15 ab$^{-1}$ and $7.3~(9.6) \times 10^{-6}$ at the FCC-hh with an integrated luminosity of 30 ab$^{-1}$, in the case the SM background is known with negligible uncertainty. When a more realistic 10\% systematic uncertainty is considered, the sensitivity decreases to $7.0~(8.5)\times 10^{-5}$ at the HE-LHC and $2.3~(3.0) \times 10^{-5}$ at the FCC-hh}}. Remarkably, then, the performance of the two machines is found roughly comparable in this case, i.e., within a factor of $\approx 3$.
Altogether, these limits are nearly two orders of magnitude better than the current  experimental results obtained from LHC runs at 13 TeV and one order of magnitude better than the existing projections for the HL-LHC at 14 TeV. Therefore, the numerical results presented here for the future HE-LHC and FCC-hh represent good reasons for pursuing further the study of their potential in extracting FCNC effects from NP manifesting themselves in top-Higgs interactions.

\vspace{0.35cm}
\noindent
{\bf Acknowledgments}\\[0.15cm]
The work of Y.-B.L is supported by the Foundation of the Henan Educational Committee (Grant no. 2015GGJS-059) and the Foundation of the Henan Institute of Science and Technology (Grant no. 2016ZD01).  SM is supported in part by the NExT Institute and the STFC CG Grant No. ST/L000296/1.

\end{document}